\documentclass[twocolumn,english,aps,prl,superscriptaddress]{revtex4}
\usepackage{graphicx}
\usepackage{amssymb}
\usepackage{hyperref}
\hypersetup{colorlinks=true,linkcolor=blue,citecolor=red,urlcolor=black}

\begin{document}

\preprint{This line only printed with preprint option}

\title{Coherent Control of Ultra-High Frequency Acoustic Resonances in Photonic Crystal Fibers}

\author{G. S. Wiederhecker}

\affiliation{Max-Planck Research Group (IOIP), University of Erlangen-Nuremberg, Guenther-Scharowsky Str. 1/Bau 24, Erlangen 91058, Germany.}

\affiliation{CePOF, Instituto de F\'{i}sica, Universidade Estadual de Campinas, 13.083-970 Campinas, SP, Brazil.}

\author{A. Brenn}

\affiliation{Max-Planck Research Group (IOIP), University of Erlangen-Nuremberg, Guenther-Scharowsky Str. 1/Bau 24, Erlangen 91058, Germany.}

\author{H. L. Fragnito}

\affiliation{CePOF, Instituto de F\'{i}sica, Universidade Estadual de Campinas, 13.083-970 Campinas, SP, Brazil.}

\author{P. St. J. Russell}

\affiliation{Max-Planck Research Group (IOIP), University of Erlangen-Nuremberg, Guenther-Scharowsky Str. 1/Bau 24, Erlangen 91058, Germany.}

\date{\today}
\begin{abstract} Ultra-high frequency acoustic resonances ($\backsim$2 GHz) trapped within the glass core ($\backsim$1 $\mu$m diameter) of a photonic crystal fiber are selectively excited through electrostriction using laser pulses of duration 100 ps and energy 500 pJ. Using precisely timed sequences of such driving pulses, we achieve coherent control of the acoustic resonances by constructive or destructive interference, demonstrating both enhancement and suppression of the vibrations. A sequence of 27 resonantly-timed pulses provides a 100-fold increase in the amplitude of the vibrational mode. The results are explained and interpreted using a semi-analytical theory, and supported by precise numerical simulations of the complex light-matter interaction. \end{abstract} \maketitle

\section{INTRODUCTION}
The interactions between optical fields and the vibrational modes of matter have long been a subject of study. Over the last two decades, it has become possible to control the strength of these vibrations by shaping the temporal (amplitude and phase) profile of an optical pulse. Demonstrations of such \textquotedblleft{}coherent control\textquotedblright{} have been reported for example for excitons in quantum wells \citep{Heberle1995}, THz lattice vibrations in crystals \citep{feurer2003}, photoisomerization of retinal molecules \citep{Prokhorenko2006} multiphoton ionization \citep{Sokolov2001} and very recently in an optical microcavity \citep{Lanzillotti07}. Along with such controlling techniques, the possibility of precisely fabricating micron-scale structures had raised renewed interest in the field due to the higher mechanical frequencies ($>$ GHz) \citep{Carmon2007-123901} and stronger susceptibility to optical forces of such tiny structures \citep{Eichenfield2007}, allowing nonlinear effects to occur at relatively low optical powers \citep{Carmon2005}. The purpose of this paper is to report coherent control of acoustic resonances (ARs) in the small solid glass core of a nonlinear photonic crystal fiber (PCF) \citep{Russell2006}. In these fibers, the cladding region surrounding the core is formed from an array of parallel air channels, see Fig. \hyperref[fig:cohcnt]{\ref{fig:cohcnt}(a,d)}. The large impedance-mismatch between glass and air creates strong confinement for acoustic waves incident upon the core-cladding interface, and in some cases a phononic band gap forms in the periodic cladding. As a result, the core (diameter $\sim$1 $\mu$m) acts essentially as a cavity for transversely-propagating acoustic waves, giving rise to trapped ARs with fundamental frequencies in the few-GHz range \citep{dainese2006}.

A short optical pulse propagating along an optical fiber excites, via electrostriction, acoustic waves \citep{dianov1990}. In standard (all-solid glass) fibers a large number of ARs are simultaneously excited, giving rise to a complicated acoustic response after propagating several km long fibers \citep{dianov1990}. In ultra-small-core PCFs the situation is dramatically different, however, because tight confinement of acoustic (and optical) waves favors excitation of only a few ultra-high-frequency ARs localized in the core. The temporal acoustic response, which can be detected after only a few dozen meters, is then the superposition of a small number of damped sinusoidal vibrations with lifetimes inversely related to the glass viscosity and acoustic leakage rate.

Here we report experiments showing that, if a second pulse is launched after the first (and well within the lifetime of the acoustic resonance), a given AR can be coherently reinforced or canceled, depending on the time-delay between the pulses. Furthermore, using a long sequence of pulses equally spaced in time, it is possible to strongly excite the particular AR whose frequency matches the repetition rate of the pulses

\section{Experiment}
In our experiments we used two different PCFs with forward Brillouin scattering spectra that had been carefully characterized \citep{dainese2006}. The first (PCF\#1, 60 m long) had a strong AR at 1.96 GHz, with a lifetime of $\tau\sim11$ ns (mechanical quality factor $Q=135$) and the second (PCF\#2, 10 m long) had one at 1.53 GHz, with a lifetime of $\tau\sim18$ ns (Q=173). Each fiber also displayed weaker ARs at higher frequencies: 2.39 GHz for PCF\#1 and 1.83 GHz for PCF\#2.
\begin{figure*}

\begin{center}

\includegraphics[width=15cm]{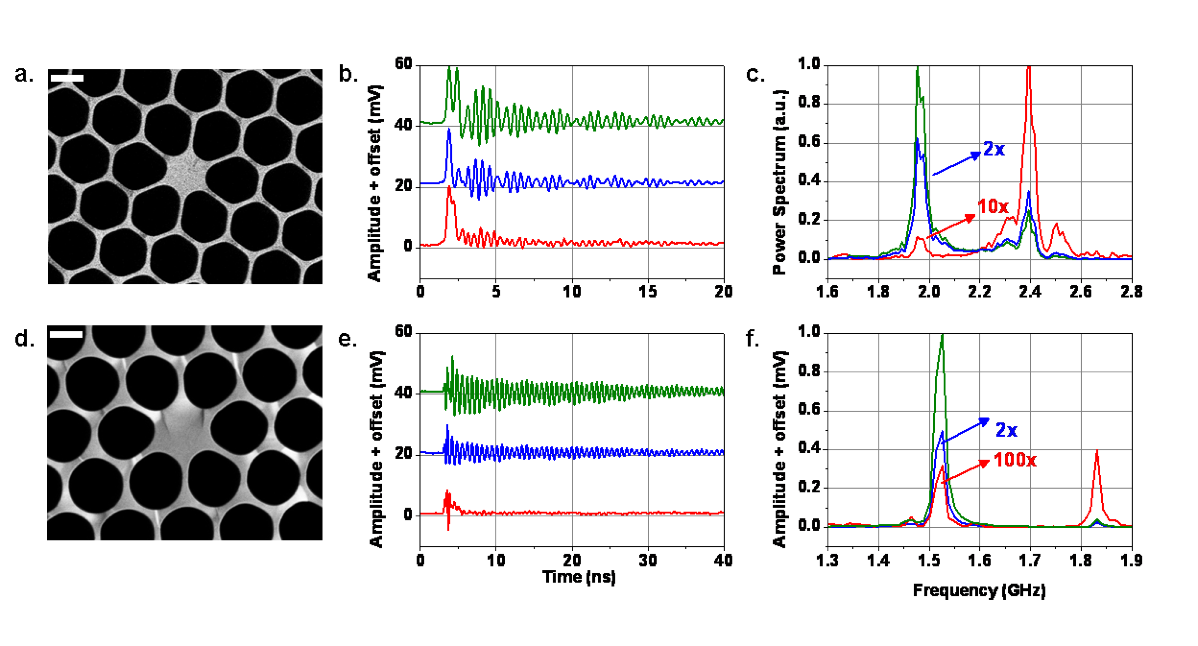}

\caption{\label{fig:cohcnt}Coherent control of acoustic phonons. The cladding pitch and core diameter are 1.67 \& 1.27 $\mu$m for PCF\#1 (tow row), and  1.92 \& 1.75 $\mu$m for PCF\#2 (bottom row). The white bars are 1 $\mu$m long. (b) Response for a single pulse excitation of PCF\#1 (blue curve), excitation by two pulses spaced by 260 ps (red curve) and excitation by two pulses spaced by 512 ps (green curve). (e) The same as b) but for PCF\#2 and pulse separations of 329 ps for the red curve and 658 ps for the green curve. The cartoons in (b) and (e) illustrate the corresponding pump pulse sequence schemes. (c)(f): The Fourier transform power spectra corresponding to the time waveforms shown in (b) and (e), corresponding curves share the same color; the arrows indicate the multiplication factor used in each curve.}

\end{center}

\end{figure*}

The experimental set-up used to excite and detect the ARs consisted of two external cavity diode lasers, a pump laser oscillating at 1530 nm and a CW probe laser at 1550 nm. An external modulator was used to create trains of 100 ps (FWHM) pulses with widely controllable inter-pulse spacings. The pulses were optically amplified, providing pulse energies of roughly 500 pJ (coupled to the PCF). After independently controlling the polarization state of the two lasers, we co-launched them into the PCF using microscope objectives. The polarization axis of the probe beam was aligned at $45\,^{\circ}$ to the PCF's birefringent axes, while the pump was polarized along its fast axis (the differential group delay between the pump and probe laser light was negligible). At the output, the pump laser light was removed using a spectral filter. A polarization-controller, followed by a polarizer, acted as our analyzer, converting the phonon-induced polarization modulation into amplitude modulation, which was then detected using a PIN photodiode and a 2.5 GHz real-time oscilloscope. The pump pulse duration (100 ps) was roughly 5 times shorter than the AR periods (510 ps and 658 ps) of the two PCFs, so that the excitation can be viewed as impulsive. Each pump pulse creates mechanical strain in the core through electrostriction \citep{Biryukov2002}, and by adjusting the delay between pulses one can induce either constructive or destructive interference in the AR.

In Fig. \hyperref[fig:cohcnt]{\ref{fig:cohcnt}(b,e)} we show oscilloscope traces of the interference between two sequentially excited ARs for PCF\#1 and PCF\#2. By adjusting the delay between the two driving pulses we were able to enhance (green curves), or suppress (red curves) the amplitude of the main AR with respect to single pulse excitation (blue curves). The oscillations observed in the red curves in Fig. \hyperref[fig:cohcnt]{\ref{fig:cohcnt}(b,e)} are caused by asynchronous ARs that are not suppressed by the second pump pulse, as well as residual oscillations of the main AR. The inital spike in all the curves is due to the fast electronic Kerr nonlinearity (cross-phase modulation) in the fibers. Figures \hyperref[fig:cohcnt]{\ref{fig:cohcnt}(c,f)} show fast Fourier transforms (FFTs) of the time-domain responses in Fig. \hyperref[fig:cohcnt]{\ref{fig:cohcnt}(b,e)}, after removing these Kerr spikes. In Fig. \hyperref[fig:cohcnt]{\ref{fig:cohcnt}c}, it is clear from the red curve that the 1.96 GHz AR has been suppressed by a factor of 30 relative to the response for a single pulse (blue curve), so that the neighboring AR at 2.39 GHz dominates the residual oscillation. PCF\#2 behaves in a very similar manner (Fig. \hyperref[fig:cohcnt]{\ref{fig:cohcnt}f}). These results show that one can carry out selective acoustic spectroscopy in such structures by preferentially exciting or suppressing targeted ARs. In Supplementary Video 1 we show how the different ARs are excited when the inter-pulse delay is scanned from 10 to 790 ps.
\begin{figure*}

\begin{center}

\includegraphics[width=15cm]{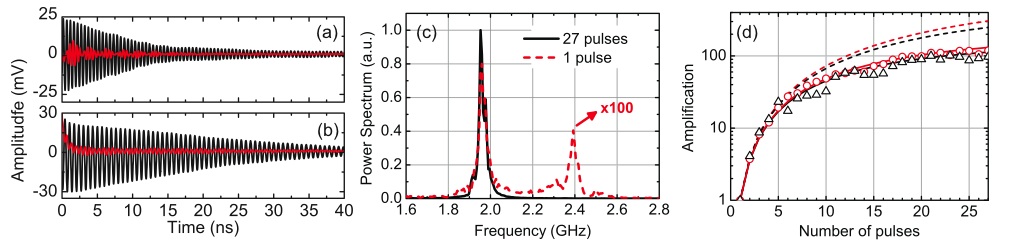}

\caption{\label{fig:mpulse}Multi-pulse excitation of ARs. (a) Response of PCF\#1 when excited by a single pulse (red curve) and a train of 27 pulses (black curve). (b) Same as (a) but for PCF \#2. (c) Fourier transform power spectra obtained from the response of PCF\#1 to a single pulse (red dashed curve) and a sequence of 27 pulses (black solid curve); the arrow indicates the multiplication factor used in the red curve. (d) Spectral power amplitude for the target AR as the number of pulses increases. The black and red curves represent PCF\#1 and PCF\#2 respectively. The dots represent the experimental data, the dashed line the expected amplification, and the solid line the expected amplification with saturation parameter $S=0.06$.}

\end{center}

\end{figure*}

To test this technique further, we generated time-sequences of pump pulses spaced by the AR period: 510 ps for PCF\#1 and 658 ps for PCF\#2. In Fig. \hyperref[fig:mpulse]{\ref{fig:mpulse}(a,b)} we compare the temporal responses of these fibers when a single pulse (red curves) and a sequence of 27 pulses (black curves) are used. To illustrate the selectivity of the process, we show in Fig. \hyperref[fig:mpulse]{\ref{fig:mpulse}c} the FFT power spectrum of PCF\#1 under these two excitation conditions. There is 100-fold enhancement in the acoustic power of the 1.96 GHz AR.

To explore how the AR amplitudes grow with pulse count, we plot on Fig. \hyperref[fig:mpulse]{\ref{fig:mpulse}d} their FFT spectral power as a function of the number of pump pulses for PCF\#1 (red circles) and PCF\#2 (black triangles). As the pulse count increases beyond 20 there is a saturation in both cases, even though the time-windows corresponding to these 20 pulses are 9.7 and 12.5 ns for PCF\#1 and \#2. One would expect the saturation to start when the number of pulses approaches $N_{sat}\approx\tau/T$ ($T$ is the inter-pulse period); this yields $N_{sat}=22$ and $N_{sat}=27$ for PCF\#1 and \#2. The expected amplification in the acoustic power spectrum for
 a sequence of $N$ pulses is $\left(1+x+...+x^{N-1}\right)^2$, with $x=e^{-T/\tau}$. This formula is plotted in Fig. \hyperref[fig:mpulse]{\ref{fig:mpulse}d} for both PCF\#1 (dashed black line) and PCF\#2 (dashed red line). In practice however, the output pulse energy from the saturated erbium-doped fiber amplifier decreases with the number $k$ of pulses in the sequence (approximately as $\left(1+kS\right)^{-1}$, where $S$ is a saturation parameter) and, therefore, the acoustic  power increases as $(\sum_{k=0}^{N-1}\frac{x^{k}}{1+kS})^2$. This formula is also plotted in Fig. \hyperref[fig:mpulse]{\ref{fig:mpulse}d} with the best fit value $S=0.06$ for both PCF\#1 (solid black line) and PCF\#2 (solid red line), in good agreement with the experimental data.

\section{Theory}
To model the observed coupling to the optical fields, one must first solve the elastic wave equation in the presence of both viscosity and electrostriction. To accomplish this we decomposed the fiber motion $\mathbf{u}$ into a superposition of all the acoustic modes $\mathbf{u}=\sum_{m}A_{m}\left|\mathbf{u}_{n}\right\rangle \exp i\Omega_{m}t$, where $\Omega_{m}$ is the modal angular frequency. Within the linear elastic approximation, one arrives at damped-driven harmonic oscillator equations for each $A_{m}$, which can be solved analytically for impulse excitation $\mathbf{E}\left(\mathbf{r},t\right)=E_{0}\delta\left(t/t_{0}\right)\left|\mathbf{\Psi}_{n}\right\rangle $, where $t_{0}$ is the pulse duration, $E_{0}$ is the field amplitude, and $\left|\mathbf{\Psi}_{n}\right\rangle$ is the normalized ($\left\langle \mathbf{\Psi}_{n}|\mathbf{\Psi}_{n}\right\rangle=1$) electric field vector corresponding to the guided optical mode ($HE_{11}$-like). The driving term is obtained by projecting the electrostrictive force on to each acoustic mode, $B_{m}=\left\langle \mathbf{u}_{m}|\mathbf{F}_{st}\right\rangle /\left\langle \mathbf{u}_{m}|\mathbf{u}_{m}\right\rangle $. The electrostrictive force density is calculated from the electric field vector, the elasto-optic constants ($p_{12}$ and $p_{44}$), and the refractive index $n$ according to $\mathbf{F}_{st}\left(\mathbf{r},t\right)=\frac{\varepsilon_{0}}{2}n^4[p_{12}\nabla\left|\mathbf{E}\right|^{2} +2p_{44}\nabla\cdot\left(\mathbf{E}\cdot\mathbf{E}^{\dagger}\right)]$ \citep{Biryukov2002}. Because of the radial symmetry of the optical mode, when pump light is coupled to one polarization state (say $\hat{\mathbf{x}}$), the force term with $p_{12}$ will point radially ($\nabla\left|\mathbf{E}\right|^{2}\propto\hat{\mathbf{r}}$) and drive only radial motion. The term with $p_{44}$, however, will point along the polarization direction ($\nabla\cdot\left(\mathbf{E}\cdot\mathbf{E}^{\dagger}\right)\propto\hat{\mathbf{x}}$) and drive a motion where the core expands in one direction while it contracts in the other.

The change in the refractive index is obtained by adding the dielectric tensor perturbation $\Delta\epsilon_{m}$, induced by each mode via the elasto-optic effect and weighted by the excitation coefficient $B_{m}$ \citep{Biryukov2002,Marcuse1975}. The refractive index modulation measured by the probe laser coupled to the mode $\left|\mathbf{\Psi}_{i}\right\rangle$ and scattered to the mode $\left|\mathbf{\Psi}_{s}\right\rangle$ is given by
\begin{equation} \Delta n\left(t\right)=\frac{U_{p}}{\rho\varepsilon_{0}cn_{eff}\left|E_{0}^{2}\right|}\sum_{m}B_{m}\left\langle \mathbf{\Psi}_{i}|\Delta\epsilon_{m}|\mathbf{\Psi}_{s}\right\rangle X_{m}\left(t\right),\label{eq:delta_n} \end{equation}
where $U_{p}$ is the pump pulse energy, $\varepsilon_{0}$ is the vacuum permittivity, $n_{eff}$ is the optical mode effective index, $c$ is the velocity of light in vacuum, and $\rho$ is the silica density. The time-dependent function is given by $X_{m}(t)=\omega_{m}^{-1}\sin(\omega_{m}t)\exp(-\Gamma_{m}t)\theta(t)$, where $\theta\left(t\right)$ is the Heaviside step-function and $\omega_{m}=\sqrt{\Omega_{m}^{2}-\Gamma_{m}^{2}}$, where $\Gamma_{m}$ is the modal damping parameter. In our experiment we are interested in inter-polarization scattering and, therefore, we choose $\left|\mathbf{\Psi}_{i}\right\rangle$ and $\left|\mathbf{\Psi}_{s}\right\rangle$ as orthogonally polarized states.
\begin{figure*}

\begin{center}

\includegraphics[width=18cm]{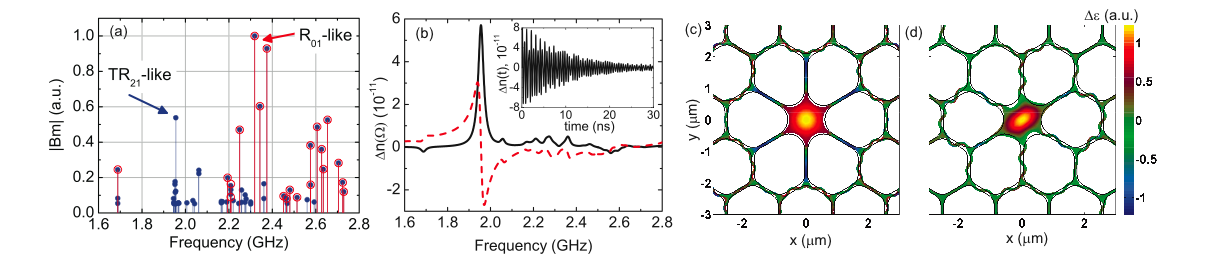}

\caption{\label{fig:calc_strict}Numerically calculated electrostriction excitation and elasto-optic refractive index modulation for PCF\#1. (a) Electrostrictive excitation for pump light polarized linearly (blue curve) and circularly (red curve); the arrows indicate the most optically active modes, shown in parts (c) and (d). (b) Real (red dashed curve) and imaginary (black solid curve) parts of the refractive index perturbation spectrum; the inset shows the temporal response. c) $\Delta\epsilon_{xx}+\Delta\epsilon_{yy}$ for the $R_{01}$-like mode at 2.318 GHz.  d) $\Delta\epsilon_{xy}$ for the $TR_{21}$-like mode at 1.956 GHz. The black solid lines represent the undeformed fiber geometry used in the calculation. The mesh deformation in the transverse plane was scaled for clarity.}

\end{center}

\end{figure*}

Focusing our attention on PCF\#1, Fig. \hyperref[fig:calc_strict]{\ref{fig:calc_strict}a} shows the numerically calculated excitation coefficient for both linear (solid blue circles) and circular (open red circles) polarization states of the pump pulses. The $TR_{2m}$-like modes are not excited with circular polarization while the $R_{0m}$-like ones are polarization insensitive. Therefore Fig. \hyperref[fig:calc_strict]{\ref{fig:calc_strict}a} is useful to identify the character of the ARs: overlapping open red and blue solid circles identify $R_{0m}$-like modes while purely blue circles identify $TR_{2m}$-like modes. Due to a dense clustering of
modes near $B_{m}=0$, we have plotted only those with  $B_{m}>5\%$. The strongest ARs in this figure (indicated by arrows) have frequencies of 1.956 GHz ($TR_{21}$-like) and 2.318 GHz
($R_{01}$-like). The profiles of displacement and dielectric tensor perturbation are shown in Fig. \hyperref[fig:calc_strict]{\ref{fig:calc_strict}(c,d)} where the radially symmetric (\hyperref[fig:calc_strict]{\ref{fig:calc_strict}c}) and torsional-radial (\hyperref[fig:calc_strict]{\ref{fig:calc_strict}d}) motion can be observed (animated versions are available in the Supplementary Videos 2 and 3). The mode amplitudes after a excitation by single 500 pJ pulse are rather small, 4 fm typically, due to the high frequency of the motions.

In Fig. \hyperref[fig:calc_strict]{\ref{fig:calc_strict}b} we show the calculated frequency dependence of the refractive index modulation (assuming $\Gamma_{m}=9.1\times10^{7}$ $s^{-1}$). The red dashed and black solid curves correspond to the real and imaginary parts. The corresponding time waveform is shown in the inset. The index modulation spectrum shows a strong peak only at 1.956 GHz ($TR_{21}$-like mode), with much weaker modulations near 2.32 GHz, in contrast to the excitation coefficients (Fig. \hyperref[fig:calc_strict]{\ref{fig:calc_strict}a}), where several modes contribute significantly. This is because most of the excited modes (large $|B_{m}|$) do not efficiently perturbs the dielectric tensor (small $|\Delta\epsilon_{m}|$). On the other hand, modes that theoretically have almost no contribution to $\Delta n$ might be observed in the experiments. This happens, for example, for the  $R_{01}$-like mode at 2.318 GHz. In a perfectly symmetric PCF (such as the one used in the simulations), the $R_{0m}$-like family behaves much like in a standard optical fiber in the sense that it does not induce any significant birefringence and therefore does not couple together orthogonally polarized optical fields. This results in a very small anisotropic index modulation for such modes. However, in real PCFs, small geometric distortions will slightly alter the exact symmetry of the $R_{0m}$ modes, allowing them to induce a degree of birefringence which can be detected in our experimental set-up. Indeed the $R_{01}$-like mode identified in Fig. \hyperref[fig:calc_strict]{\ref{fig:calc_strict}a} oscillates at 2.318 GHz, which is close to the experimentally-measured AR at 2.39 GHz.

In conclusions, ARs trapped in the small core of highly nonlinear PCFs, oscillating at GHz frequencies, can be coherently controlled by sequences of optical driving pulses with durations shorter than a half-cycle of acoustic oscillation. Such multi-pulse excitation can be used to preferentially amplify or attenuate selected resonances. Control of ultra-high frequency resonances may have applications in mode-locking of fiber lasers at very high repetition rates, in the realization of optically-pumped acoustic \textquotedblleft{}sasers\textquotedblright{}, or in the design of more efficient fiber-based acousto-optic devices.

We would like to acknowledge N. Joly for fabricating the fibers, and the financial support from the Brazilian agencies CAPES, CNPq, and FAPESP.

\bibliographystyle{apsrev}
\bibliography{references_rev2}

\begin{thebibliography}{13}
\expandafter\ifx\csname natexlab\endcsname\relax\def\natexlab#1{#1}\fi
\expandafter\ifx\csname bibnamefont\endcsname\relax
  \def\bibnamefont#1{#1}\fi
\expandafter\ifx\csname bibfnamefont\endcsname\relax
  \def\bibfnamefont#1{#1}\fi
\expandafter\ifx\csname citenamefont\endcsname\relax
  \def\citenamefont#1{#1}\fi
\expandafter\ifx\csname url\endcsname\relax
  \def\url#1{\texttt{#1}}\fi
\expandafter\ifx\csname urlprefix\endcsname\relax\def\urlprefix{URL }\fi
\providecommand{\bibinfo}[2]{#2}
\providecommand{\eprint}[2][]{\url{#2}}

\bibitem[{\citenamefont{Heberle et~al.}(1995)\citenamefont{Heberle, Baumberg,
  and Kohler}}]{Heberle1995}
\bibinfo{author}{\bibfnamefont{A.~P.} \bibnamefont{Heberle}},
  \bibinfo{author}{\bibfnamefont{J.~J.} \bibnamefont{Baumberg}},
  \bibnamefont{and} \bibinfo{author}{\bibfnamefont{K.}~\bibnamefont{Kohler}},
  \bibinfo{journal}{\prl} \textbf{\bibinfo{volume}{75}}, \bibinfo{pages}{2598}
  (\bibinfo{year}{1995}).

\bibitem[{\citenamefont{Feurer et~al.}(2003)\citenamefont{Feurer, Vaughan, and
  Nelson}}]{feurer2003}
\bibinfo{author}{\bibfnamefont{T.}~\bibnamefont{Feurer}},
  \bibinfo{author}{\bibfnamefont{J.~C.} \bibnamefont{Vaughan}},
  \bibnamefont{and} \bibinfo{author}{\bibfnamefont{K.~A.}
  \bibnamefont{Nelson}}, \bibinfo{journal}{Science}
  \textbf{\bibinfo{volume}{299}}, \bibinfo{pages}{374} (\bibinfo{year}{2003}).

\bibitem[{\citenamefont{Prokhorenko et~al.}(2006)}]{Prokhorenko2006}
\bibinfo{author}{\bibfnamefont{V.~I.} \bibnamefont{Prokhorenko}}
  \bibnamefont{\textit{et~al.}}, \bibinfo{journal}{Science}
  \textbf{\bibinfo{volume}{313}}, \bibinfo{pages}{1257} (\bibinfo{year}{2006}).

\bibitem[{\citenamefont{Sokolov et~al.}(2001)}]{Sokolov2001}
\bibinfo{author}{\bibfnamefont{A.~V.} \bibnamefont{Sokolov}}
  \bibnamefont{\textit{et~al.}}, \bibinfo{journal}{\prl} \textbf{\bibinfo{volume}{8703}}
  (\bibinfo{year}{2001}).

\bibitem[{\citenamefont{Lanzillotti-Kimura et~al.}(2007)}]{Lanzillotti07}
\bibinfo{author}{\bibfnamefont{N.~D.} \bibnamefont{Lanzillotti-Kimura}}
  \bibnamefont{\textit{et~al.}}, \bibinfo{journal}{\prl} \textbf{\bibinfo{volume}{99}},
  \bibinfo{pages}{217405} (\bibinfo{year}{2007}).

\bibitem[{\citenamefont{Carmon and Vahala}(2007)}]{Carmon2007-123901}
\bibinfo{author}{\bibfnamefont{T.}~\bibnamefont{Carmon}} \bibnamefont{and}
  \bibinfo{author}{\bibfnamefont{K.~J.} \bibnamefont{Vahala}},
  \bibinfo{journal}{\prl} \textbf{\bibinfo{volume}{98}},
  \bibinfo{pages}{123901} (\bibinfo{year}{2007}).

\bibitem[{\citenamefont{Eichenfield et~al.}(2007)}]{Eichenfield2007}
\bibinfo{author}{\bibfnamefont{M.}~\bibnamefont{Eichenfield}}
  \bibnamefont{\textit{et~al.}}, \bibinfo{journal}{Nat. Photonics}
  \textbf{\bibinfo{volume}{1}}, \bibinfo{pages}{416} (\bibinfo{year}{2007}).

\bibitem[{\citenamefont{Carmon et~al.}(2005)}]{Carmon2005}
\bibinfo{author}{\bibfnamefont{T.}~\bibnamefont{Carmon}} \bibnamefont{\textit{et~al.}},
  \bibinfo{journal}{\prl} \textbf{\bibinfo{volume}{94}},
  \bibinfo{pages}{223902} (\bibinfo{year}{2005}).

\bibitem[{\citenamefont{Russell}(2006)}]{Russell2006}
\bibinfo{author}{\bibfnamefont{P.~St.J} \bibnamefont{Russell}},
  \bibinfo{journal}{J. Lightwave Technol.} \textbf{\bibinfo{volume}{24}},
  \bibinfo{pages}{4729} (\bibinfo{year}{2006}).

\bibitem[{\citenamefont{Dainese et~al.}(2006)}]{dainese2006}
\bibinfo{author}{\bibfnamefont{P.}~\bibnamefont{Dainese}} \bibnamefont{\textit{et~al.}},
  \bibinfo{journal}{Opt. Express} \textbf{\bibinfo{volume}{14}},
  \bibinfo{pages}{4141} (\bibinfo{year}{2006}).

\bibitem[{\citenamefont{Dianov et~al.}(1990)}]{dianov1990}
\bibinfo{author}{\bibfnamefont{E.~M.} \bibnamefont{Dianov}}
  \bibnamefont{\textit{et~al.}}, \bibinfo{journal}{Opt. Lett.}
  \textbf{\bibinfo{volume}{15}}, \bibinfo{pages}{314} (\bibinfo{year}{1990}).

\bibitem[{\citenamefont{Biryukov et~al.}(2002)\citenamefont{Biryukov, Sukharev,
  and Dianov}}]{Biryukov2002}
\bibinfo{author}{\bibfnamefont{A.~S.} \bibnamefont{Biryukov}},
  \bibinfo{author}{\bibfnamefont{M.~E.} \bibnamefont{Sukharev}},
  \bibnamefont{and} \bibinfo{author}{\bibfnamefont{E.~M.}
  \bibnamefont{Dianov}}, \bibinfo{journal}{Quantum Electronics}
  \textbf{\bibinfo{volume}{32}}, \bibinfo{pages}{765} (\bibinfo{year}{2002}).

\bibitem[{\citenamefont{Marcuse}(1975)}]{Marcuse1975}
\bibinfo{author}{\bibfnamefont{D.}~\bibnamefont{Marcuse}},
  \bibinfo{journal}{Bell Syst. Tech. J.} pp. \bibinfo{pages}{985--995}
  (\bibinfo{year}{1975}).

\end{thebibliography}

\end{document}